%% file: main.tex
\documentclass[journal]{IEEEtran} % use the `journal` option for ITherm conference style
\IEEEoverridecommandlockouts
\usepackage{cite}
\usepackage{amsmath,amssymb,amsfonts}
\usepackage{graphicx}
\usepackage{textcomp}
\usepackage{xcolor}
\usepackage{svg}

\usepackage{hyperref}
\usepackage{float}
\usepackage{tikz}
\usepackage{circuitikz}
\usepackage[sort&compress,numbers]{natbib}
\usepackage{amsmath}
\usepackage{fontawesome5}
\usepackage{rotating}
\usepackage{pifont}

\usepackage{algorithm}
\usepackage[noend]{algpseudocode}
\usepackage{tabularx}
\usepackage{xltabular}
\usepackage{longtable}
\usepackage{makecell}

\usepackage{lineno}
\usepackage{todonotes}

\makeatletter
\def\BState{\State\hskip-\ALG@thistlm}
\makeatother

\makeatletter
\def\BState{\State\hskip-\ALG@thistlm}
\makeatother

\usepackage{helvet}
\usepackage{tabularray}

\usetikzlibrary{shapes}
\usetikzlibrary{arrows}

\tikzstyle{every task} = []
\tikzstyle{every gateway} = []
\tikzstyle{every sequence} = []
\tikzstyle{every message} = []
\tikzstyle{every association} = []
\tikzstyle{every event} = []

\tikzstyle{task} = [rectangle, draw, black,
minimum width=4em, minimum height=2em,rounded corners,align=center,every task]

\tikzstyle{gateway} = [diamond, draw, black,inner sep=0pt,minimum width=2.5em, minimum height=2.5em,every gateway]
\tikzstyle{sequence} = [->,>=triangle 45,every sequence]
\tikzstyle{message} = [o->,dashed,>=open triangle 45,every sequence]
\tikzstyle{association} = [->,densely dotted,>=angle 45,every association]
\tikzstyle{event} = [circle,minimum width=1.5em, minimum height=1.5em,draw,every event]
\tikzstyle{end event} = [event,ultra thick,every event]
\tikzstyle{intermediate event} = [event,double,every event]

\usetikzlibrary{arrows.meta, automata,
	calc,
	positioning,
	quotes,
	fit}

\tikzstyle{mybox} = [draw,rectangle, minimum height=0.75cm,minimum width=4cm]

\newcommand{\vertices}{V}
\newcommand{\edges}{E}
\newcommand{\executables}{T}

\newcommand{\hcurr}{h_{current}}
\newcommand{\hnew}{h_{new}}

\newcommand{\scurr}{s_{current}}
\newcommand{\snew}{s_{new}}

\newcommand{\Scurr}{S_{current}}
\newcommand{\Snew}{S_{new}}

\newcommand{\rcurr}{r_{current}}
\newcommand{\rnew}{r_{new}}

\newcommand{\Ccurr}{C^{enc}_{curr}}
\newcommand{\Cnew}{C^{enc}_{new}}

\newcommand{\sk}{sk}
\newcommand{\pk}{pk}

\def\BibTeX{{\rm B\kern-.05em{\sc i\kern-.025em b}\kern-.08em
    T\kern-.1667em\lower.7ex\hbox{E}\kern-.125emX}}
\begin{document}

\title{Blockchain-Based, Confidentiality-Preserving Orchestration of Collaborative Workflows}
% delete or comment-out the following line before submission

\author{%%%% author names
    \IEEEauthorblockN{Bal\'{a}zs \'{A}d\'{a}m Toldi}% first author
    , \IEEEauthorblockN{Imre Kocsis}% delete this line if not needed
    %, \IEEEauthorblockN{3\textsuperscript{rd} Given Name Surname}% delete this line if not needed
    % duplicate the line above as many times as needed to list all authors
    \\%%%% author affiliations
    \IEEEauthorblockA{\textit{Dept. of Measurement and Information Systems\\ Budapest University of Technology and Economics\\Budapest, Hungary}}\\% first affiliation
    %\IEEEauthorblockA{\textit{Department of Measurement and Information Systems --- Budapest University of Technology and Economics}}\\% delete this line if not needed
    % duplicate the line above as many times as needed to list all affiliations
    %%%% corresponding author contact details
    \IEEEauthorblockA{balazs.toldi@edu.bme.hu, kocsis.imre@vik.bme.hu}
}

\maketitle

\input{content/abstract.tex}

\input{content/introduction.tex}

\input{content/background.tex}

\input{content/aproach.tex}

\input{content/BPMN.tex}

\input{content/mapping}

\input{content/protocol}

\input{content/implementation}
%\input{content/securityGuarantees.tex}

%

%\input{content/DnO.tex}

%\input{content/testing.tex}

%\section{Implementation?}

\input{content/results.tex}

\input{content/limitations.tex}

\input{content/conclusion.tex}

\section*{Acknowledgment}
This work was partially created under, and financed through, the Cooperation Agreement between the Hungarian National Bank (MNB) and the Budapest University of Technology and Economics (BME).
%The preferred spelling of the word ``acknowledgment'' in America is without
%an ``e'' after the ``g''. Avoid the stilted expression ``one of us (R. B.
%G.) thanks $\ldots$''. Instead, try ``R. B. G. thanks$\ldots$''. Put sponsor
%acknowledgments in the unnumbered footnote on the first page.

%\section*{References}

%Please number citations consecutively within brackets \cite{b1}. The
%sentence punctuation follows the bracket \cite{b2}. Refer simply to the reference
%number, as in \cite{b3}---do not use ``Ref. \cite{b3}'' or ``reference \cite{b3}'' except at
%the beginning of a sentence: ``Reference \cite{b3} was the first $\ldots$''
%
%Number footnotes separately in superscripts. Place the actual footnote at
%the bottom of the column in which it was cited. Do not put footnotes in the
%abstract or reference list. Use letters for table footnotes.
%
%Unless there are six authors or more give all authors' names; do not use
%``et al.''. Papers that have not been published, even if they have been
%submitted for publication, should be cited as ``unpublished'' \cite{b4}. Papers
%that have been accepted for publication should be cited as ``in press'' \cite{b5}.
%Capitalize only the first word in a paper title, except for proper nouns and
%element symbols.
%
%For papers published in translation journals, please give the English
%citation first, followed by the original foreign-language citation \cite{b6}.

\bibliography{refs/mybib}

\begin{IEEEbiography}[{\includegraphics[width=1in,height=1.25in,clip,keepaspectratio]{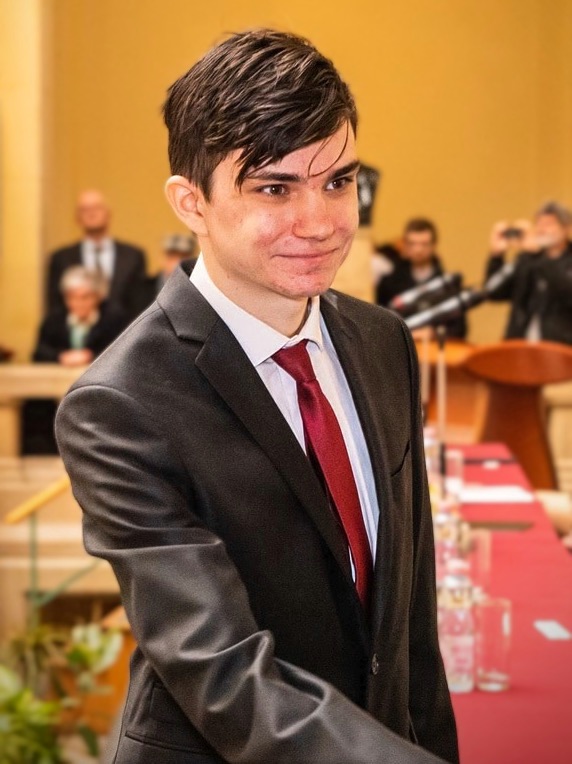}}]{Balázs Ádám Toldi}
received  his BSc in computer engineering in 2023 from the Budapest University of Technology and Economics (BME). Currently, he is an MSc student at BME, with a primary specialization in cybersecurity and a secondary specialization in critical systems.
\end{IEEEbiography}
\begin{IEEEbiography}[{\includegraphics[width=1in,height=1.25in,clip,keepaspectratio]{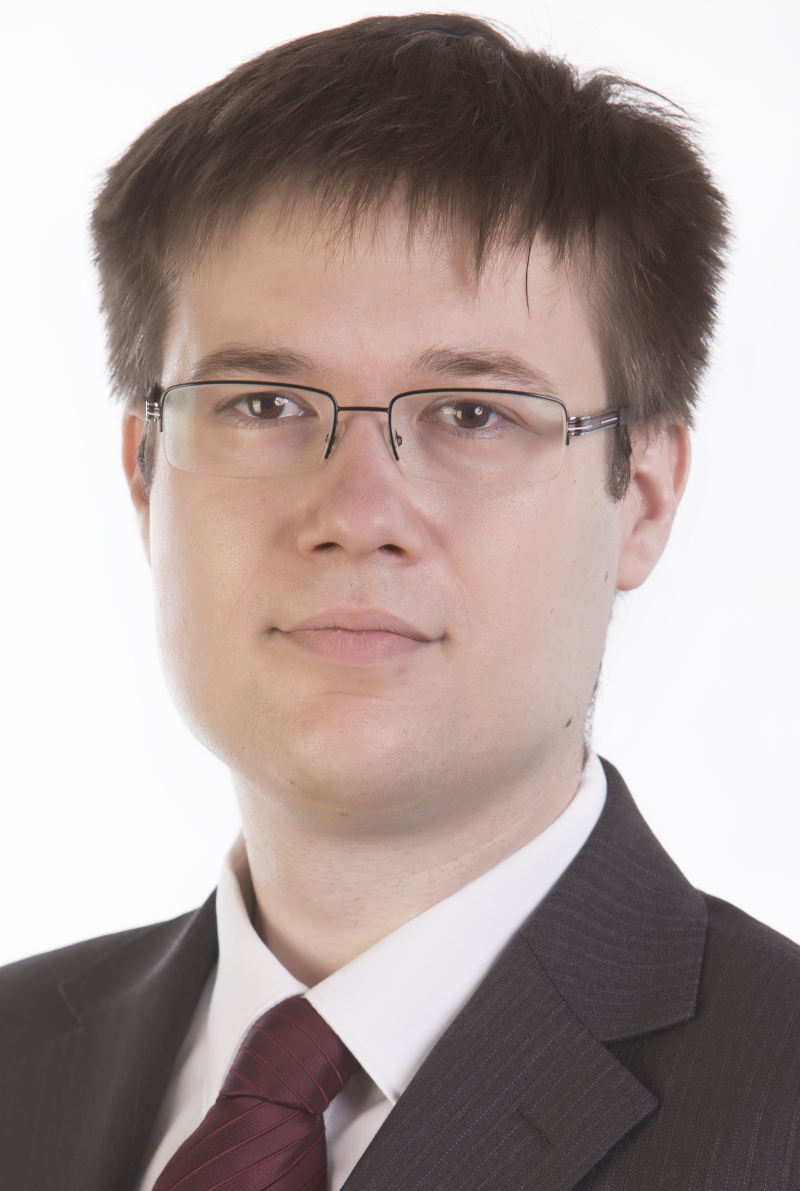}}]{Imre Kocsis} received his PhD from the Budapest University of Technology and Economics (BME) in 2019. Currently, he serves as a senior lecturer and leading blockchain researcher at the Critical Systems research group of the Dept. of Measurement and Information Systems of BME. He leads the activities of the group in conjunction with the Hyperledger Foundation and the university's participation in the European Blockchain Services Infrastructure (EBSI) network.
\end{IEEEbiography}
%\begin{thebibliography}{00}
%    \bibitem{b1} G. Eason, B. Noble, and I. N. Sneddon, ``On certain integrals of Lipschitz-Hankel type involving products of Bessel functions,'' Phil. %Trans. Roy. Soc. London, vol. A247, pp. 529--551, April 1955.
%    \bibitem{b2} J. Clerk Maxwell, A Treatise on Electricity and Magnetism, 3rd ed., vol. 2. Oxford: Clarendon, 1892, pp.68--73.
%    \bibitem{b3} I. S. Jacobs and C. P. Bean, ``Fine particles, thin films and exchange anisotropy,'' in Magnetism, vol. III, G. T. Rado and H. Suhl, Eds. %New York: Academic, 1963, pp. 271--350.
%    \bibitem{b4} K. Elissa, ``Title of paper if known,'' unpublished.
%    \bibitem{b5} R. Nicole, ``Title of paper with only first word capitalized,'' J. Name Stand. Abbrev., in press.
%    \bibitem{b6} Y. Yorozu, M. Hirano, K. Oka, and Y. Tagawa, ``Electron spectroscopy studies on magneto-optical media and plastic substrate interface,'' %IEEE Transl. J. Magn. Japan, vol. 2, pp. 740--741, August 1987 [Digests 9th Annual Conf. Magnetics Japan, p. 301, 1982].
%    \bibitem{b7} M. Young, The Technical Writer's Handbook. Mill Valley, CA: University Science, 1989.
%\end{thebibliography}
\vspace{12pt}

%\section{Remaining TODO}
%\begin{itemize}
%    \item "Nemgagyi" video
%    \item Table I cleanup
%    \item Fix Table I: \url{https://tex.stackexchange.com/questions/152067/tikz-picture-in-table}
%    \item Specify nature of randomness
%    \item Modellt lelevelezni
%    \item Publishers for the references
%    \item Checking the title casing for the references
%\end{itemize}

\end{document}

%% file: content/abstract.tex
\begin{abstract}
Business process collaboration between independent parties is challenging when participants do not completely trust each other. Tracking actions and enforcing the activity authorizations of participants via blockchain-hosted smart contracts is an emerging solution to this lack of trust, with most state-of-the-art approaches generating the orchestrating smart contract logic from Business Process Model and Notation (BPMN) models. However, compared to centralized business process orchestration services, smart contract state typically leaks potentially sensitive information about the state of the collaboration, limiting the applicability of decentralized process orchestration. This paper presents a novel, collaboration confidentiality-preserving approach where the process orchestrator smart contract only stores encrypted and hashed process states and validates participant actions against a BPMN model using zero-knowledge proofs. We cover a subset of BPMN, which is sufficient from the practical point of view, support message-passing between participants, and provide an open-source, end-to-end prototype implementation that automatically generates the key software artifacts. %Under our approach, no party external to the collaboration can gain trustable knowledge of the current state of a process instance (barring collusion with a participant), even if it has full access to the blockchain history.
\end{abstract}

\begin{IEEEkeywords}
blockchain, BPMN, orchestration, collaboration, confidentiality, zero-knowledge proofs
\end{IEEEkeywords}

%% file: content/introduction.tex
\section{Introduction}
%\linenumbers
In modern business science, \textit{Business Process Management} (BPM) as a discipline~\cite{van2016business} advocates process-focused thinking about internal activities and external collaborations to improve key performance indicators. Automating the execution of business processes is a key proposition of BPM and has been supported for a long time by various technical solutions~\cite{POURMIRZA201743}. Today, most of these, typically centralized, tools and services use the leading business process modeling standard, Business Process Model and Notation (BPMN) 2.0~\cite{bpmn} as a process definition language~\cite{CHINOSI2012124}.

Distributed ledger technology (DLT), generally implemented on a blockchain basis, is widely recognized as a compelling platform to support the cross-organisational execution of business processes -- even when the organisations cannot agree on a trusted (third) party as a middleman~\cite{10.1145/3183367}. Blockchain-deployed smart contracts can impartially enforce the agreed-on sequences of activities and track sent and received messages. Smart contracts can also host data objects acted on by a process directly or anchor their changes in the blockchain via cryptographic commitments. 

However, blockchain-assisted BPM is still a relatively new discipline -- importantly, known BPMN-based solutions are inadequate from the privacy and confidentiality point of view. This paper presents a novel, collaboration confidentiality-preserving approach and end-to-end prototype tooling for the on-chain process orchestration of cross-organizational, BPMN-based collaborations using zero-knowledge proofs (ZKPs)\footnote{This paper is based on the Scientific Student Association report submitted by Balázs Ádám Toldi to the 2022 competition at the Budapest University of Technology and Economics: \url{https://tdk.bme.hu/VIK/sw8/Kollaborativ-munkafolyamatok-titkossagmegorzo}}. Specifically, for a sufficient subset of BPMN, we present a transformation of the admissible state updates of BPMN process instances to programs of the ZoKrates~\cite{zokrates} toolkit. We assemble state update validity provers from these programs for the participants and proof-verifying orchestrator smart contracts. We define an on-chain process state commitment update protocol, describe our open-source end-to-end implementation prototype\footnote{Available at \url{https://github.com/ftsrg/zkWF}} and evaluate practical viability. %The current implementation supports Ethereum-based blockchains and Hyperledger Fabric and is easy to extend for further platforms.

Our contribution is novel from two aspects. First, to our knowledge, the confidentiality challenges of decentralized BPMN orchestration have not been addressed systematically and constructively yet. Second, we express BPMN execution as an incremental computation in a form amenable to commit-and-prove style zero-knowledge validation in smart contracts. This paves the way for further research on the computational representation of orchestrated BPMN execution against the continuously appearing ZKP advancements.

%\textbf{TBD The paper is structured as follows.}

\section{Motivation and problem statement}
\label{sec:motiv}
BPMN is a standardized approach to visually and precisely express \textit{how} business processes should be performed. BPMN is used in many domains -- including finance, banking, manufacturing, healthcare, logistics and telecommunications -- for capturing processes with well-defined sequences of regularly repeated activities. The BPMN standard defines several model types, \textit{process}, \textit{collaboration} and \textit{choreography} being the most widely used ones. \textit{Process (flow)} models are the simplest: these express the sequence, preconditions and exception handling of a single process performed by a single organization. Collaborations model the individual processes performed by collaborating parties -- usually business entities -- and their messaging-based interactions. Choreography diagrams focus solely on the message exchanges between collaborating entities.

%with message exchange flows and activity sequence coordination.

 %Such tools usually include rich technical support for process automation, e.g., to generate web pages for performing activities or to integrate external activity implementations.

\subsection{Decentralized orchestration}
For over a decade, software tools have been available to assist with \textit{process execution}. The more sophisticated ones track and \textit{orchestrate} activities according to a BPMN model, register activity-related data and perform decision-making on further process evolution. However, centralized orchestration introduces a trusted orchestrator party requirement when we move beyond single-entity processes. With the emergence of blockchain and distributed ledger technology, the potential of decentralizing various aspects of cross-organizational collaboration has been recognized quite early.

%in general, and the orchestration of BPMN-expressed collaborations in particular,

\begin{figure*}
\centering
\includegraphics[width=0.95\textwidth]{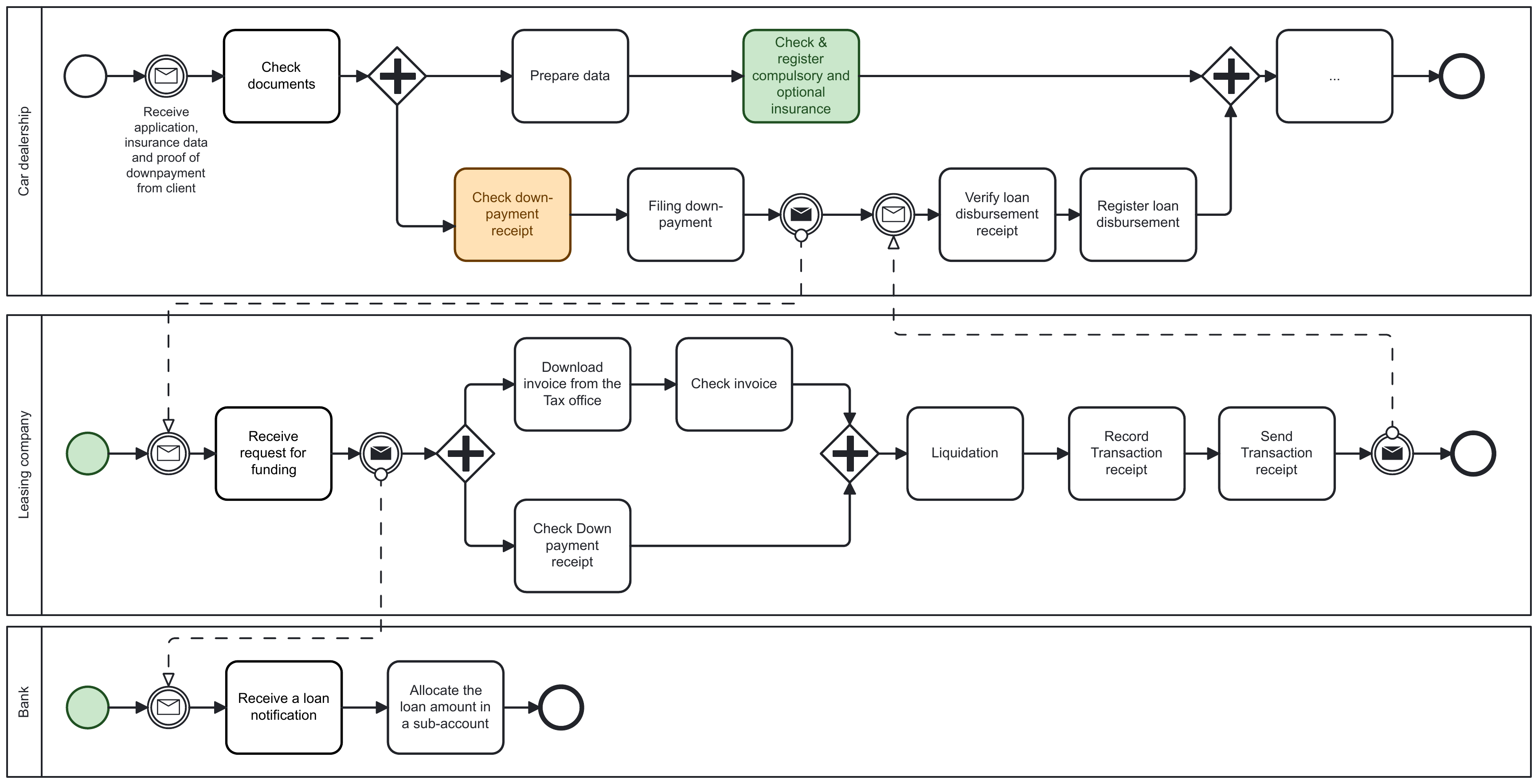}
\caption{Car leasing BPMN collaboration example (simplified for presentation). In the depicted example state, green denotes ''completed''; orange is ''active''.}
\label{fig:example}
\end{figure*}

Consider the BPMN car leasing collaboration model in Figure~\ref{fig:example}\footnote{The model was created in the ''Digitisation, artificial intelligence and data age workgroup'' of the ongoing BME-MNB cooperation project. (MNB is the Central Bank of Hungary.). For legibility, the process in Figure~\ref{fig:example} is slightly simplified; the whole model is available in our project repository.}, where the internal processes of a car dealership, a leasing company and a financing bank must be coordinated to accept a leasing application. Individual executions of models are called \textit{instances} and have an \textit{instance state}. The coloring in Figure \ref{fig:example} demonstrates a state: green denotes that the dealership has completed insurance processing and the leasing company and the bank would be able to begin processing. However, for the leasing company to proceed, active downpayment checking (orange) must be finished, then the downpayment filed, and a downpayment notification sent and received.

Using blockchain-deployed smart contracts that track \textit{collaboration state}, the \textit{execution enablement} and \textit{execution obligation} of the activities of the parties can be enforced without a dedicated, trusted party. The transaction journal nature of blockchains can also ensure that the full trace is also stored in an immutable and irrepudiable way. While tracking the internal state of participant-internal processes on-chain is not always desirable, it is a valuable \textit{option}; e.g., when decisions have to be made in a way verifiable by the other collaborating parties.

Orchestrating and journaling messages and collaborative data handling are two further collaboration aspects which can be improved with ''blockchainification''. %These are a) facilitating message passing and securing the messages themselves; and b) reading and writing the data used collaboratively by the processes (expressed via ''data objects'' and ''data stores'' in BPMN). In both cases, mainly for performance and cost reasons, the dominant approach is to use smart contracts to store cryptographic (hash) commitments to externally handled messages and data modifications.
%(expressed via ''data objects'' and ''data stores'' in BPMN). 
%In both cases, to avoid storing sizeable data on-chain, the technical approach is usually to manage cryptographic (hash) commitments to externally handled messages and data modifications in the orchestrator smart contracts.
In both cases, the orchestrator smart contracts usually only manage cryptographic (hash) commitments to externally handled messages and data modifications, to avoid storing sizeable data on-chain.

Tools and approaches exist to create orchestrator smart contracts from BPMN models (see Section~\ref{sec:procorch}). However, no systematic solution exists to protect sensitive collaboration state information in the smart contract state from parties who can read the blockchain but do not participate in the collaboration. In our example, a leasing company may wish that its competitors do not see how many open cases they have, how long it takes to perform key steps in the process, or what lease rates they apply.

Fulfilling such requirements is a confidentiality challenge that contradicts core blockchain design principles. Blockchain nodes must be able to \textit{validate} and \textit{execute} incoming transaction requests to reach consensus on ledger updates, be those changes of the balances of a natively tracked cryptocurrency or state changes of deployed smart contracts. If the transaction details are made ''incomprehensible'' to the nodes, e.g., by off-chain encryption, they can't validate the preconditions for performing the transaction and compute state updates. %In this case, smart contracts can still serve as resilient storage for encrypted state updates or hash commitments but lose the ability to validate the request \textit{semantically}; in our case, whether the right party proposes the right state-advancement of the workflow. 
%For smart contracts, the two dominant, general-purpose cryptographic approaches to the privacy/confidentiality versus transaction validation/executability conundrum are validating transactions with ZKPs and confidentiality-preserving execution using homomorphic encryption, with the prior being significantly better established currently.
For smart contracts, the dominant \textit{cryptographic} answers to this dilemma are validating transactions with ZKPs and confidentiality-preserving execution using homomorphic encryption, with the prior being significantly better established currently.

\subsection{Problem statement: BPMN collaboration confidentiality}
\label{subsec:prob}
We set up our problem statement through a basic system model and the enumeration of required security properties. We target a simple form of collaboration confidentiality (see the properties below) under the assumption that it is not in the interest of any process participant to leak information about process instances; participants neither directly leak information nor help external parties to compromise confidentiality. This is one of the realistic models for our setting, even though the participants do not completely trust the \textit{actions} of each other. We will touch briefly on stronger models in Section~\ref{sec:futur}.

%However, these have not yet been applied to the on-chain orchestration of BPMN-defined collaborations; we present an approach using the non-interactive ZKP route.

\subsubsection{Basic assumptions and terminology} \textit{participants} wish to collaborate in the execution of an \textit{instance} of a previously agreed-on BPMN collaboration definition. All other parties are \textit{process external}. All participants have a cryptographic key pair for signature-based authentication and process activity authorisation. The underlying process model is public knowledge, but the public keys are shared only between the participants. We assume the absence of private key compromises. 

%These key pairs are ''application-level'' and independent of those used for basic transaction signing at the blockchain platform level.

For the underlying blockchain, we assume complete integrity (no successful attack on the consensus) and, for the sake of simplicity, deterministic finality (accepted blocks do not get retracted). Note that even blockchains with probabilistic block finality are usually quasi-deterministically final already at the time scale of a few blocks. On the other hand, process external parties have complete visibility of blockchain transactions. We treat the blockchain as \textit{fair} -- any transaction submitted by a participant is included in a block in a reasonable time, irrespective of concurrent transaction request load. While, in practice, blockchain platforms have strongly varying fault and threat models and sensitivity (see, e.g., \cite{cachin2017blockchain}), these are basic assumptions of normal operational conditions. As a part of platform selection, security and dependability analysis should evaluate the risk of these assumptions not being met.

\subsubsection{System model} the classic Business Process Orchestrator (BPO) middleware pattern \cite{gorton2006essential} facilitates business process execution by providing a message broker and extending it with state management and persistent state storage. The solutions in the state-of-the-art closely match this pattern. (Technically, message passing is only \textit{coordinated} and journaled by the smart contract their core.) The smart contract as a Process Controller \cite{gorton2006essential} also performs authentication and authorization based on the BPMN model to ensure that the stored state sequence never deviates from the model semantics. We also aim to employ a blockchain-deployed smart contract as a BPO.

%through which participants asynchronously track their respective progress in the process, exchange messages and which serves as a trustworthy and resilient log of the execution trace. 

\subsubsection{Security properties} we target a set of integrity, availability and confidentiality guarantees. Integrity and availability properties are already covered by the prior art; our contributions lie in establishing collaboration confidentiality, as defined by properties \textbf{C1} and \textbf{C2}, despite using smart contracts. BPO-SC refers to a per-process instance BPO smart contract.

\smallskip

\noindent \textbf{I1}: The state traces enforced by the BPO-SC always adhere to the operational semantics of the underlying BPMN model.

\smallskip

\noindent \textbf{I2}: Process-external parties cannot influence the BPO-SC state.

\smallskip

\noindent \textbf{I3}: Process state updates can be initiated only by the participants authorized by the model and instance state.

\smallskip

\noindent \textbf{A1}: No external party can influence authorised participants' ability to perform state updates in a bounded time.

\smallskip

\noindent \textbf{A2}: No participant can influence the ability of authorized participants to perform state updates in a bounded time.

\smallskip

\noindent \textbf{A3}: Participants can always learn the trace and current state.

\smallskip

\noindent \textbf{C1}: External parties cannot determine participant identities.

\smallskip

\noindent \textbf{C2}: No external party can learn more about the trajectory, timings and stepwise properties (e.g., process variables and message contents) of the trace during and after execution than the fact that an instance has been started.

%% file: content/background.tex
\section{Related work}
\label{sec:procorch}
Smart contracts, as a rule, cannot be altered after deployment; thus, to minimize the probability of software faults, domain-specific languages and Model-Driven Engineering (MDE) are steadily gaining ground in smart contract development~\cite{model_driven}. In our context, the established approach is a BPMN model to serve as a \textit{specification}, and orchestrator smart contract logic is generated automatically from the model.

%In our own review of the state of the art, three tools emerged as the most mature and most relevant to our research: Caterpillar, Lorikeet and Chorchain. These provide valuable templates for future research, but in the end, due to the core difference from our own approach, We could not reuse them.
\subsection{Decentralized business process orchestration}
Caterpillar~\cite{caterpillar} was the first open-source BPMN-to-Solidity compiler (Solidity is the primary smart contract development language for the Ethereum platform). %Its primary purpose is to execute collaborative business processes between mutually distrusting parties on blockchains. 
Since its initial release, several forks have emerged. Some of these also come with an extended feature set, like Blockchain Studio~\cite{Mercenne_Brousmiche_Hamida_2018}, which adds role management, or~\cite{Abid_Cheikhrouhou_Jmaiel_2020}, which adds time constraints. Lorikeet~\cite{lorikeet} is a model-driven engineering approach that integrates assets into business processes. Lorikeet extends the BPMN 2.0 specification with support for asset registries and also transforms models into Solidity smart contracts. The smart contracts handle the orchestration of the process as well as interactions with the tokens. Chorchain~\cite{chorchain} takes a BPMN \textit{choreography} and generates an Ethereum smart contract that can be used to execute the model. ChorChain also includes a dedicated modeling tool. The same authors released two further tools: Multi-Chain~\cite{multichain} and FlexChain~\cite{flexchain}. Multi-chain is similar to Chorchain, but it also supports Hyperledger Fabric~\cite{hyperledger_fabric}. FlexChain can only produce Solidity smart contracts, but the user can also define a ruleset for each choreography. If a condition in the ruleset is met, then an off-chain processor will perform its underlying action. 

Our analysis showed that the process state and trace are easily recoverable from the process manager smart contracts for \textit{all} the tools above.

%Beyond BPMN, the \href{https://docs.baseline-protocol.org/}{Baseline protocol}\footnote{\url{https://docs.baseline-protocol.org/}} is a developing open standard that allows enterprises to synchronize complex, multi-party business processes on distributed ledger technologies. Business process workflows in Baseline are formed as state machines. The standard includes some essential and optional privacy-related requirements. The protocol has two reference implementations; however, at the time of this writing, neither of these actually supports privacy/confidentiality measures.

\subsection{Commit-and-prove ZKP with smart contracts}
Zero-Knowledge Proofs (ZKPs) are cryptographic methods to prove the validity of various statements without revealing any additional information~\cite{ZKProofCommunity}. ZKP verification in a smart contract requires a scheme with ''single-shot'' message passing from prover to verifier; in this work, we rely on zk-SNARKs, a family of \textit{noninteractive}, and also \textit{succinct} (small and cheaply verifiable proofs) ZKPs. We use the ZoKrates toolkit as a ZKP front-end with a high-level programming language~\cite{zokrates}. ZoKrates currently supports the Groth16~\cite{Groth_2016}, GM17~\cite{Groth_Maller_2017} and Marlin~\cite{cryptoeprint:2019/1047} proving schemes.

Our contribution implements a commit-and-prove approach. In commit-and-prove schemes, a party first commits to an input and, possibly later, proves some predicate about the input -- without revealing it~\cite{zkws4}. This is a widely used pattern in the smart contract-based application of zero-knowledge proofs. Recent surveys on ZKP schemes, technologies and applications can be found in \cite{9520375} and~\cite{9300214}.

%% file: content/aproach.tex
\section{A confidentiality-preserving approach}
\label{sec:conf}
The fundamental difference of our approach from \cite{caterpillar,chorchain,multichain,flexchain,lorikeet,Mercenne_Brousmiche_Hamida_2018,Abid_Cheikhrouhou_Jmaiel_2020} is that instead of storing the process state on-chain in an easily interpretable form in an orchestrator smart contract, we store encrypted states and cryptographic state commitments and accept update proposals on the presentation of ZKPs over the current and proposed commitment. The approach relies on two key conceptual components: our zkWF (''zero knowledge WorkFlow'') protocol and what we call ''zkWF programs''.

\subsection{The zero-knowledge WorkFlow (zkWF) protocol}
\label{subsec:prot}
\begin{figure}
\centering
\includegraphics[width=0.75\columnwidth]{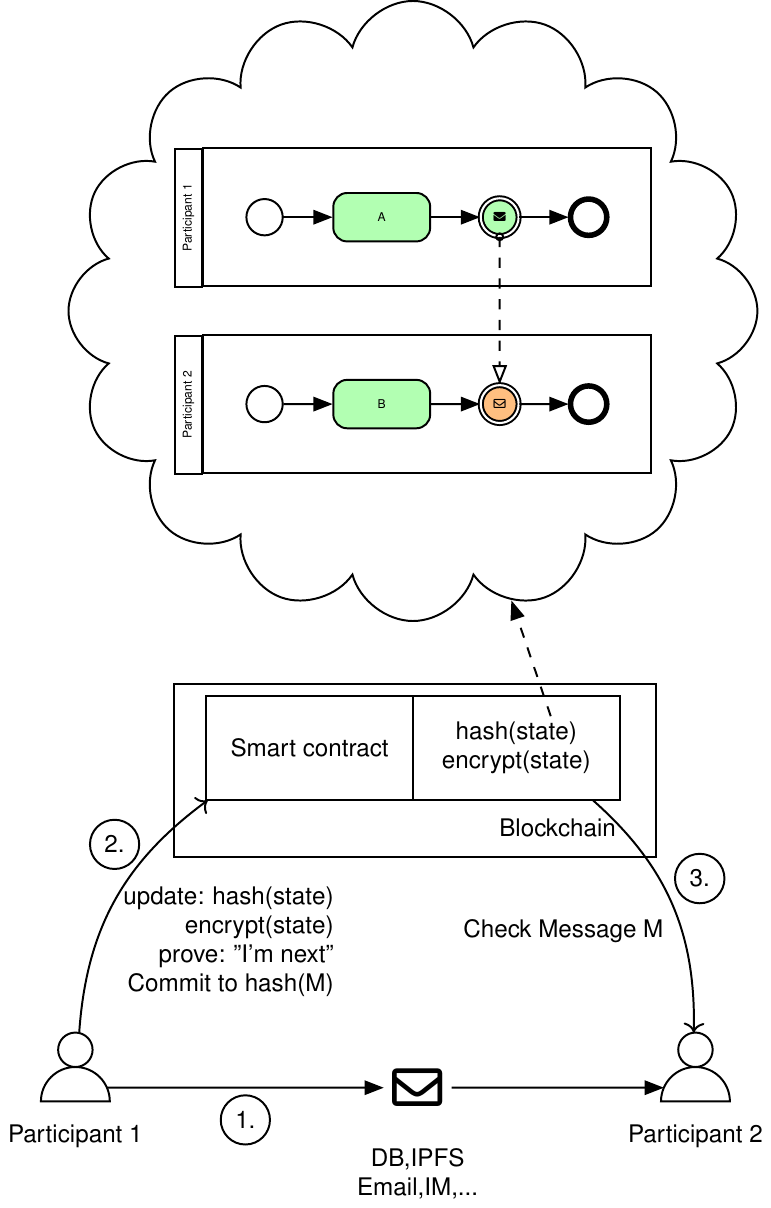}
    \caption{Overview of the zkWF protocol}
    \label{fig:basic_update}
\end{figure}

The zkWF protocol is a hash commitment style protocol that allows the participants of a business process to follow and step the execution of a business process. Figure \ref{fig:basic_update} presents a high-level overview.

At the centre of the scheme is a smart contract instance on a blockchain for an instance of a collaboration model. This smart contract stores and manages the state of the collaboration -- as specified by the underlying BPMN model -- in an encrypted and a hashed form.

During process execution, the collaborating parties can send messages to each other by off-chain means \ding{172}. These are captured in the underlying process specification as intermediate message throw and capture events; our state commitment scheme includes commitments to the message hashes. 

When a participant wishes to update the state stored in the smart contract -- that is, to ''step the process'' --, it has to create a ZKP that the proposed state transition is valid. This new state includes the hash of the message they sent beforehand if the step involves message sending. It sends the new state hash commitment, the encrypted new state and the ZKP proof of state transition validity to the smart contract as a blockchain transaction \ding{173}; the smart contract updates its state only if it can successfully check the ZKP.

When the execution arrives at a point where a participant receives a message in the next stage of the execution, the receiving party checks the hash and only accepts (and proceeds with its part of the collaboration) if the hashes match \ding{174}.

Participant authentication is tied to proving private key ownership in the ZKPs. The public keys are defined over the participant group-shared process model as a parameterization. These are cooperation-private, ''application-level'' key pairs; on pseudonymizing platforms, such as Ethereum, updater identity can and should be masked by using independent, single-use transaction source addresses (i.e., public keys).
%On pseudonymizing platforms, such as Ethereum, it also facilitates masking updater identity as single-use transaction source addresses (i.e., public keys) can be used independently of the cooperation-private keypair used ''at the application level''.

Additionally, we require the participants to have a common means for encrypting and decrypting stored state ciphertexts. The protocol does not constrain the encryption used.

The protocol can be realized straightforwardly on a wide range of DLTs; we provide an implementation for Ethereum and Hyperledger Fabric ~\cite{hyperledger_fabric}.  While the updates and the contract state are unintelligible to parties outside the collaboration, statistical and model trace analyses of the update sequences are still a threat. We enable mitigations by including a ''fake'' update transaction variant (no actual state update), which all participants are authorized to use.

\subsection{zkWF programs}
zkWF programs are generated from BPMN specifications and serve as a bridge between process definition and proof computation/verification. A zkWF program is a ZoKrates program that, for a given BPMN model instance (parameterized model), can decide whether a given actor is authorized to execute a state transition in a given execution state. We use the zkWF program to generate the zero-knowledge proofs and proof verification code for the orchestrator smart contract.

\subsection{Workflow and toolchain}
We created an end-to-end toolchain prototype for our approach, as depicted in Figure~\ref{fig:vezerabra}. %The figure also identifies the newly created software components and the artefacts with novel generators.

In the \textit{modeling phase}, a BPMN model is annotated with metadata for process instantiation, and our interpreter-translator creates the corresponding zkWF program.

In the \textit{synthesis phase}, the ZoKrates toolkit is used to set up the \textit{prover key} and \textit{verifier key} and generates the verifier smart contract in Solidity. We created novel support for generating verifier code for Hyperledger Fabric in Java. We also created the code generation facilities for both platforms' state commitment management part of the smart contracts.

Some secret values used when creating zk-SNARK prover and verifier keys are considered ''toxic waste'': an adversary can use them to break the scheme, e.g., forge fake proofs. Thus, security relies on the waste having been deleted. The associated risk can be mitigated by using a reliable party for the key generation or performing so-called multi-party trusted setup ceremonies, where a (large) group of actors assembles the keys. In this case, security requires only at least one of them to delete the waste. Such ceremonies tend to be complicated and thus can pose a problem for by-program setup. Universal schemes also exist (e.g.,~\cite{cryptoeprint:2019/1047}), where the results of a single program-agnostic ceremony can be used to derive program-specific keys publicly and securely. Choosing the right approach requires deployment-specific risk analysis; ZoKrates supports all of the above.
%, and consequently, our implementation does, too. We do note that currently, we require a setup for each process instance.

For the \textit{deployment phase}, we created automation facilities for deployment to Ethereum (and other blockchains using a compatible RPC API); and an SDK and GUI for the client side. Here, we integrate the ZoKrates toolkit as a proof generator.

\noindent
\input{content/tikz-figure/vezer}

%is key-authorized process state changes to be possible only in terms of the smart contract state commitment and stored encrypted state. 
%Note that this goal carries over from the existing state of the art.

%Our \textit{availability goal} is no process external party to stop

%In this paper, we assume the following properties may apply for a given attacker:

%\begin{itemize}
%\item An attacker knows the corresponding BPMN model for a given smart contract instance
%\item A malicious participant may also be an attacker
%\end{itemize}

%Our approach aims at ensuring the following security guarantees.
%\todo{Ezeket befelyezni}
%\begin{itemize}  
% \item Parties not participating cannot modify the state stored on the blockchain
% \item Parties not participating cannot read or guess the current state of the business process execution based on the data stored in the smart contract. 
%\item No party can make an illegal move during the orchestration.
%\item Messages between the participants can be verified.
%\item Participants can identify which party made a given move 
%\end{itemize}

%% file: content/tikz-figure/vezer.tex
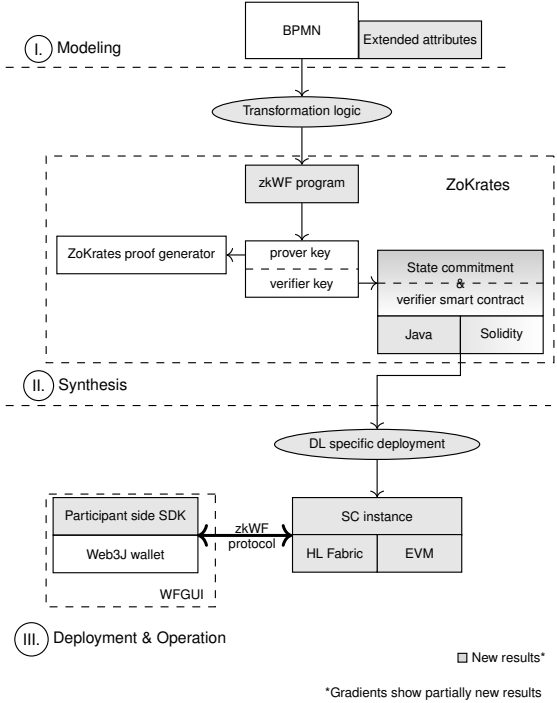
\begin{figure}
	\centering
	\begin{tikzpicture}[font=\sffamily,line width=0.015cm, scale=0.5,every node/.style={scale=0.5}]
	    \node[rectangle,draw,minimum width= 3cm,minimum height=1.5cm](BPMN){BPMN};
	    \node[rectangle,draw,minimum width= 3cm,minimum height=1cm,right=0 cm of BPMN,xshift=\pgflinewidth,yshift=-0.25cm,fill=gray!20!white,xshift=-\pgflinewidth](ext){Extended attributes};
	    \node[below=0.3 cm of BPMN,xshift=-8cm,yshift=0.5cm] (m1) {};
	    \node[below=0.3 cm of ext,xshift=2.5cm,yshift=0.5cm] (m2) {};
	    \draw[dashed] (m1) -- (m2);
	    \node[above=0.5cm of m1,xshift=1cm,yshift=-1cm,draw,circle](t1){\large{I.}};
	    \node[right=0.5cm of t1,xshift=-1cm]{\large{Modeling}};
	    
	    \node[below=0.5cm of BPMN,draw,ellipse,yshift=0cm,fill=gray!20!white] (transform) {Transformation logic};
	    \draw[<-] (transform.north) -- (BPMN.south);
	    \node[below=0.5cm of transform,draw,rectangle,yshift=-0.0cm,minimum width= 3cm,minimum height=1cm,fill=gray!20!white] (zkwf) {zkWF program};
	    
	    \node[right=1.1cm of zkwf] (zokrates) {\large ZoKrates};
	    
	    \draw[<-] (zkwf.north) -- (transform.south);
	    \node[below=0.5 cm of zkwf,draw,rectangle,yshift=-0.0cm,minimum width= 3cm,minimum height=1.5cm,align=left] (keys) {prover key\\\\ verifier key};
	    \draw[<-] (keys.north) -- (zkwf.south);
	    \draw[dashed] (keys.west) -- (keys.east);
	    \node[left=0.25 cm of keys,draw,rectangle,minimum width=4.5cm,minimum height=1cm,yshift=0.375cm] (proofGen){ZoKrates proof generator};
	    \node[left=0.5cm of keys.west,xshift=1.25cm,yshift=0.375cm] (proverKeyBorder){};
	    \draw[<-] (proofGen.east) -- (proverKeyBorder);
	    \node[right=0.5 cm of keys.east,xshift=-1.25cm,yshift=-0.375cm] (verifierKeyBorder){};
	    \node[right=0.25 cm of keys,align=center,draw,rectangle,minimum width= 4.4cm,minimum height=1.75cm,yshift=-0.375cm,top color=gray!40!white, bottom color=white] (sc) {State commitment\\\&\\verifier smart contract};
	    \draw[<-] (sc.west) -- (verifierKeyBorder);
	    \draw[dashed] (sc.west) -- (sc.east);
	    \node[below=0.5cm of sc,draw,rectangle,yshift=1.0cm,minimum width=2.2cm,minimum height=1cm,xshift=-1.1cm,fill=gray!20!white,yshift=\pgflinewidth] {Java};
	    \node[below=0.5cm of sc,draw,rectangle,yshift=1.0cm,minimum width=2.2cm,minimum height=1cm,xshift=1.1cm,left color=gray!20!white, right color=white,yshift=\pgflinewidth] (solc){Solidity};
	    
	    \node[draw,rectangle,dashed, fit=(zkwf) (solc) (proofGen),minimum width=13.5cm,minimum height=5.5cm]{};
	    
	    \node[below=1.375cm of keys,xshift=-8cm] (s1) {};
	    \node[below=1.375cm of keys,xshift=6.75cm] (s2) {};
	    \draw[dashed] (s1) -- (s2);
	    \node[above=0.5cm of s1,xshift=1cm,yshift=-1cm,draw,circle](t2){\large{II.}};
	    \node[right=0.5cm of t2,xshift=-1cm]{\large{Synthesis}};

	    \node[below=1.75cm of keys,draw,ellipse,fill=gray!20!white,xshift=2cm] (depl) {DL specific deployment};
	    \node[below=0.1cm of sc.south,yshift=0.25cm,xshift=0.0cm] (scBot){};
	    \draw[->] (scBot) -| ++(0,-1.5) -| (depl.north); 
	    
	    \node[below=0.5cm of depl,rectangle,draw,minimum width=4.5cm,minimum height=1cm,fill=gray!20!white](sci){SC instance};
	    \draw[<-] (sci.north) -- (depl.south);
	    \node[below=0 cm of sci,rectangle,draw,minimum width=2.25cm,minimum height=1cm,yshift=\pgflinewidth,xshift=1.125cm,fill=gray!20!white,yshift=\pgflinewidth]{EVM};
	    \node[below=0 cm of sci,rectangle,draw,minimum width=2.25cm,minimum height=1cm,yshift=\pgflinewidth,xshift=-1.125cm,fill=gray!20!white,yshift=\pgflinewidth]{HL Fabric};
	    
	    \node[left=0.5cm of sci,draw,rectangle,xshift=-1.5cm,minimum width=3.85cm,minimum height=1cm,fill=gray!20!white](sdk){Participant side SDK};
	    \node[below=0cm of sdk,rectangle,draw,minimum width=3.85cm,minimum height=1cm,yshift=\pgflinewidth](web3){Web3J wallet};
	    \node[left=0.5cm of sci.south,xshift=-1cm] (sc_side){};
	    \node[right=0.5of sdk.south,xshift=0.655cm] (sdk_side){};
	    \draw[<->,line width=0.04cm] (sc_side) -- (sdk_side);
	    \node[left=0.5cm of sc_side,xshift=0.5cm,yshift=0.2cm,align=center](zkwfp){zkWF};
	    \node[below=0.0cm of zkwfp,yshift=0.1cm]{protocol};
	    \node[below=0.5 cmof web3,yshift=0.7cm,xshift=1.5cm](GUI) {WFGUI};
	    
	    \node[draw,dashed,fit=(sdk) (GUI) (web3),minimum width=4.5cm, minimum height=3cm]{};
	    
	    \node[below=0.5cm of sci,yshift=-2cm,xshift=3.5cm](note){New results*};
	    \node[draw,rectangle,fill=gray!30!white,left=0.5cm of note,xshift=1cm]{};
	    \node[below=0.2cm of note,xshift=-2cm]{*Gradients show partially new results};
	    
	    \node[below=0.5 cm of GUI,xshift=-4cm,yshift=0.5cm,draw,circle](t3){\large{III.}};
	    \node[right=0.5 cm of t3,xshift=-1cm]{\large{Deployment \& Operation}};
	\end{tikzpicture}
	\caption{Toolchain overview}
	 \label{fig:vezerabra}
\end{figure}

%% file: content/BPMN.tex
\section{BPMN subset and execution semantics}
%\todo{Intro, subset, extensions}
This paper targets the Basic Modeling Elements of BPMN 2.0 \cite[p.~28]{bpmn}, the core subset of the specification, with the restrictions that regarding events, we interpret only message throw and catch ones (among participants) and do not support sub-processes and data objects. We argue that this element set is already sufficient for practical applications. Statistical evidence~\cite{Muehlen2013} shows that the usage frequency of the 50 constructs in the BPMN specification follows a Zipfian distribution; we cover elements used at least in $\sim25\%$ of the models in~\cite{Muehlen2013}. This is also the empirically established ''Common Core of BPMN'' in~\cite{Muehlen2013} with the addition of messaging between participants. Our earlier example showcases the supported element set (except for ''exclusive gateways'' for process variable-based choice paths and ''lanes'' for further subdividing pools).

%Currently, we support a limited but representative set of elements from the BPMN specification, as summarized by table \ref{tab:sup_elem}. Notably, in addition to the Basic Modeling Elements of BPMN 2.0 (see \cite{bpmn}, p28), we also support message throw and catch events, which are of particular importance in collaborative settings. The table denotes those elements as ''stateful'' which have non-instantaneous execution semantics (as declared by the BPMN specification), and these will determine the structure of our execution state vector. In the context of this paper, we will refer to these elements as the ''executable'' ones in the BPMN subset we address.

%Some of the element types are considered executable. Others are there to control the execution flow. Executable elements' (like that of activities) state is tracked by this tool. The currently supported elements can be seen in table \ref{tab:sup_elem}.

%We wanted to support at least the Basic Modeling Elements of BPMN 2.0\footnote{See Business Process Model and Notation, v2.0 \cite{bpmn}, page 28.}  to prioritize common components. We have also taken into account making modeling collaborations more executable. This is why We also included Message flows and Intermediate Message events. We chose these elements because these are the ones that are necessary and sufficient to model most of the relevant use cases.

\subsection{BPMN extensions and structural constraints}
%\label{bpmn_attr}
We introduce two extended attributes for BPMN elements. \texttt{zkp:publicKey} separates the tasks of different participants by attaching a participant-specific public key to a pool, a lane, or a \textit{participant-executable element} (activities, message throws and catches). %Applying this attribute to an element directly or indirectly (e.g. through inclusion in a pool) is mandatory; the intended usage is to equip either pools or lanes with public keys. 
\texttt{zkp:variables} applies to \textit{activities} and declares process instance global variables, and that that activity may write the variable (reads are allowed for all activities). These variables can be used in boolean expressions for exclusive gateways.

Some constraints apply to the structure of the BPMN models, which are currently admissible in our scheme.

\begin{itemize}
    \item Gateways must be binary (two incoming/outgoing edges).
    \item Activities are \textit{atomic}; i.e., subprocesses are not supported.
    \item The model must be acyclic (no loops).
\end{itemize}

We plan to eliminate these constraints in the future; the required modifications of the state representation and the zkWF program construction are largely incremental.

\subsection{State representation}
\label{subsec:staterep}
Our notion of process instance execution state encompasses the following aspects (for the specific encoding in zkWF programs, please refer to the report and the implementation).

\begin{itemize}
\item A vector $v$ of the current state of executable elements
\item The current values of \textit{global variables}
\item Hashes of the messages already sent in the process
\end{itemize}

Let $M=(\vertices,\edges,\executables)$ be a process model, where $\vertices$ is the set of non-flow model elements, $\edges$ is the set of model edges (flows), and $\executables\subset\vertices$ is the set of all executable elements in the business process. Then, $v$ is a vector of $|T|$ size and $\forall v_i \in v$ can have one of the following three values:

\begin{itemize}
    \item 0 (Inactive) -- The element has not been reached yet
    \item 1 (Active) -- The element is ready to be executed or is being executed by a participant
    \item 2 (Completed) -- The execution of the element has been completed
\end{itemize}

This state set is a subset of those in the standard activity lifecycle \cite[p.~428]{bpmn} and serves as a reasonable simplification, as the main focus of the work described here is exploring the confidential execution aspect. Note that correctly implementing the full lifecycle is a significant software engineering effort, even in the centralized setting. Also, BPMN users tend to apply a similar simplified view during modeling, as the more sophisticated state aspects require experience and limit the ease of model understanding. %The major exception is the ''Ready'' state between Inactive and Active, which we folded into our Active state due to not being required under our system model -- there seems to be limited value in a participant declaring that she has started performing a task.

\subsection{Capturing token passing semantics}
\label{subsec:parray}
BPMN 2.0 models have straightforward, token flow-based standard execution semantics: start events create tokens that move around as execution progresses. Parallel gateways split and join tokens. To support a different ZKP use case, \cite{toots_msc} introduces a technique for representing valid BPMN execution state changes by enumerating the possible composite token marking deltas of the elements upon stepping the process. Specifically, \cite{toots_msc} introduces an array $P$, where each element of $P$ is a list of token change and element identifier pairs. We construct a similar $P$ array under the token passing semantics and embed it into the zkWF program to enable checking whether a proposed state update is valid from the BPMN execution logic point of view. Our $P$ array to describe one-step token marking changes for a model $M$ consists of 3-tuples with elements from the set $\mathcal{N}$: 

\begin{align}
    \mathcal{N}&=(+1, -1\}\times T)\cup \{(0,-1)\}
\end{align}

For $T$, we apply a simple integer encoding; the $-1$ in the ''no-token-change'' pair second set is a don't care placeholder. Especially under our binary gateway condition, which is currently necessary to ensure reasonable proof computation times, it is straightforward to enumerate the admissible changes based on the BPMN model. For example, let's consider activities $a,b,c \in T$. $a$ continues in a parallel gateway, which proceeds to $b$ and $c$. When $a$ transitions from ''Active'' to ''Completed'' and $b$ and $c$ from ''Inactive'' to ''Active'', the following token marking change happens: $((-1,a),(+1,b),(+1,c)) \in \mathcal{N}$. The complete logic can be found in the referenced report.

%% file: content/mapping.tex
\section{zkWF program and protocol design}
\label{sec:design}
%We first introduce our mapping logic of BPMN models to zkWF programs. As zkWF programs represent process \textit{instances}, we also describe our state representation approach. This is one of the key contributions of our work; the state commitment scheme of the zkWF protocol is built on this basis.

%\subsection{zkWF program design}
%The zkWF program construction is a central contribution of this paper. It is generated for each BPMN model and used to generate zero-knowledge proofs.

A zkWF program is a ZoKrates program shared among the participants, with which process participants prove that a business process state transition they propose is allowed. In ZKP terms, the participants are the \textit{provers}, and the orchestrator smart contract is the \textit{verifier}. %The smart contract stores cryptographic state commitments and an encrypted version of the state and manages their updates.

ZoKrates programs have public as well as private inputs, and an output. Private inputs are only visible to the prover; public inputs are visible to the prover and the verifier, and they are necessary to verify proofs. In our case, the current commitment and the proposed one act as public inputs. Private inputs are more varied; only some are shared across the participants (e.g., the cleartext of the current state).%, and some are specific to the individual participants.

%zkWF programs have public as well as private inputs, and an output. Private inputs are only visible to the respective prover. Public inputs are visible to the prover as well as the verifier, and they are necessary to verify proofs.  %Outputs are similar to public inputs, but the user does not supply these; they are the results of the executed program with some private and public inputs.

%In general terms, the computation happens against a current public process state commitment, stored by the managing smart contract, which is the hash of the current process state, salted by some randomness. This ensures that parties outside the collaboration can't easily guess states from their hash commitments. The computation also relies on the knowledge of the current state and the randomness used for salting the commitment of the current state -- these are shared between the collaborating parties in an encrypted form through the smart contract. 

The key current deficiency of our scheme is that our proofs do not include showing the congruence of the on-chain stored state ciphertexts and the public state (hash) commitments. Combining established encryption algorithms with zk-SNARKs is hard; advances are being made (see, e.g., \cite{cryptoeprint:2019/1270}), but these haven't appeared in any of the leading zk-SNARK frameworks yet as vetted and reusable ''gadgets''.%, in stark contrast to other primitives, e.g., hash and signature algorithms.

We apply the following measures to this deficiency. An additional part of our public input (and blockchain-stored data) will be a \textit{signature commitment}: the current hash commitment and the \textit{previous} hash commitment signed by the last acting party (using their application-level cryptographic identity). Should a participant erroneously or maliciously commit a ciphertext that does not hash to the stated, proven and accepted commitment, this signature ensures that the offending participant can be irrepudiably identified by the other collaborating parties. 

Although several partially mitigative and corrective schemes can be built on this measure, we introduce the weakening assumption that the irrepudiable identifiability of participants halting execution this way is a sufficient disincentive.

\subsection{zkWF computation model}
\begin{figure}
\centering
\includegraphics[width=0.7\columnwidth]{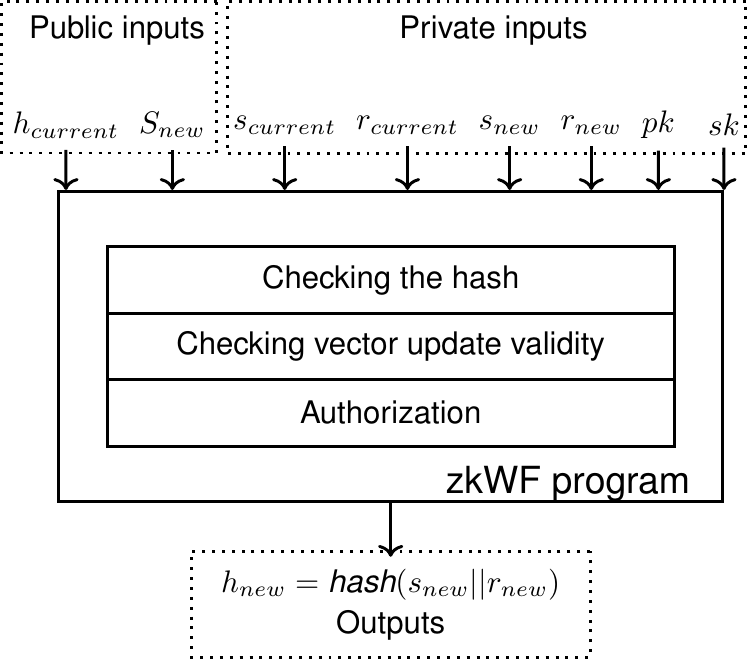}
\caption{The basic computation model of zkWF programs}
\label{fig:zkwfProgram}
\end{figure}

Figure \ref{fig:zkwfProgram} illustrates the structure of the generated zkWF programs. For hashing, we use SHA-256; application-level signing uses the EdDSA implementation from the ZoKrates standard library (both widely used, NIST-standard algorithms). The \textit{private} inputs of zkWF programs are as follows.

\begin{itemize}
	\item $\scurr$ - the current state of the process (subsec. \ref{subsec:staterep})
	\item $\rcurr$ - random salt for hashing $\scurr$ (32 bits)
	\item ${\snew}$ - the updated (''stepped'') process state
	\item $\rnew$ - new randomness, for hashing $\snew$
	\item $\pk$ - public EdDSA key of the participant (subsec. \ref{subsec:prot})
	\item $\sk$ - private EdDSA key of the participant
\end{itemize}

\smallskip

\noindent The \textit{public} inputs ($\Vert$ denotes concatenation):
\begin{itemize}
	\item $\hcurr =\textit{hash}(\scurr || \rcurr)$
        \item $\Snew=\textit{sig}(\hcurr||\hnew)$	
        %\item $\Snew$ - ($\hcurr \Vert \hnew$), signed by the currently\todo{Cross-check, hogy nem ostobaság-e a last acting helyett currently actinget mondani.} acting participant.
\end{itemize}

\smallskip

\noindent $sig$ denotes signing by the party proposing the new hash commitment in the concatenation. Given these inputs, the following steps are performed.

\begin{enumerate}
    \item Checking the group-shared secret current state and randomness against the public hash commitment to ensure ongoing integrity. %The zkWF program also checks the signature on the concatenation of the current and proposed commitments.
    \item Checking that no illegal state transition is being proposed through $\snew$ at the \textit{process logic} level.
    \item Checking the new signature commitment given as a public input (based on $pk$ and $sk$) %, to ensure that the participant has the correct key pair, to be signed by the acting participant 
    and checking the authorization of the participant for the business process step.
    \item The program outputs the hash of the new state.
\end{enumerate}

Most aspects of the computational model are straightforward; we only expand on the important details of BPMN model encoding and the state change validity checking logic.

\subsection{BPMN model encoding and state change validation}
\label{subsec:encode}
The BPMN model logic is carried over into the zkWF program by a precomputed $P$ array (Section \ref{subsec:parray}). To check whether the correct paths are proposed for exclusive gateways, the expressions on the sequence flows after the gateways are also encoded in the program as assertions. Message passing and variable write permission checks are addressed similarly.

%We have introduced the inputs and outputs of the zkWF programs; described the core approach for encoding BPMN models; and specified the ZKP framework to use. 

%In this section, we discuss the main steps of zkWF programs, as also identified on Figure \ref{fig:zkwfProgram}. 

%\subsubsection{Checking the supplied hash}
%\label{checking_hash}
%The hash of the current state and its corresponding randomness must be the hash present in the smart contract. This step ensures the integrity of the business process execution.

%\subsubsection{Proving that a state vector update is valid}
%\label{method}
%This proving process step ensures that a proposed step is valid in the BPMN model.

Regarding the executable element state vector, the program compares $v_{current}$ and $v_{new}$ from $s_{current}$ and $s_{new}$. If the two are the same, the ''change'' is accepted (as a ''step'' under our fake update mechanism). Four or more differences (pairwise comparisons at the same indices) in the vectors are considered invalid. Otherwise, we construct a $3\times3$ matrix $A$ with the initial value

\begin{align}
A = \begin{bmatrix}
		0 & -1 \\ 0 & -1 \\ 0 & -1
\end{bmatrix}
\end{align}

Then, for the $j$-th difference ($j \in 0\ldots 2$) at position $i \in 0 \ldots |T|-1$ in the vectors, we apply the following updates:

\begin{itemize} 
        \item $v_{current}[i] = 1 \And v_{new}[i] = 2 \Rightarrow A[j] \leftarrow [-1,i]$
        \item $v_{current}[i] = 0 \And v_{new}[i] = 1 \Rightarrow A[j] \leftarrow [1,i]$
        \item $v_{current}[i] = 0 \And v_{new}[i] = 2 \Rightarrow A[j] \leftarrow [1,i]$
\end{itemize}

Any other combination of $v_{current}$ and $v_{new}$ values is invalid. If $P$ contains an element with the rows of $A$, then token passing-wise, the proposed state change is valid, as we essentially decoded the activity token marking changes ($\pm1$) from the activity state changes: 0 - Inactive $\rightarrow$ 1 - Active $\rightarrow$ 2 - Completed.

%It constructs a new matrix $A$ in the following way:
%\begin{enumerate}
%	\item At first,take matrix $A$ as $\begin{bmatrix}
%		0 & -1 \\ 0 & -1 \\ 0 & -1
%	\end{bmatrix}$ (no changes) and $j=0$ as a counter
%	\item Compare every $v_{old}[i]$ and $v_{new}[i]$, where $i\in [0,T[$ 
%	\begin{itemize} 
%			\item If $v_{old}[i] = 1$ and $v_{new}[i] = 2$, replace $A[j]$ with $[-1,i]$
%			\item If $v_{old}[i] = 0$ and $v_{new}[i] = 1$, replace $A[j]$ with $[1,i]$
%			\item If $v_{old}[i] = 0$ and $v_{new}[i] = 2$, replace $A[j]$ with $[1,i]$ 
%			\item If none of the above are true, but $v_{old}[i] \neq v_{new}[i]$ , replace $A[j]$ with $[-1,-1]$ (invalid change) 
%			\item If $v_{old}[i] \neq v_{new}[i]$ then increase $j$ by one
%		\end{itemize}
%\end{enumerate}

%Zero changes are also considered valid. This makes it easy to generate "fake" state changes: the process state in the smart-contract changes, but in reality, the state vector does not. This can be useful to mask the current state of the process execution. %See section \ref{hashing_reason} for more details.

%Four or more changes in the process state are considered invalid. % The reason behind it is described in section \ref{paralhell}.

%\paragraph{Problem with parallel gateway ends}
Parallel gateway ends (''joins'') induce an additional check: a transition from a state where \textit{not both activities before the gateway are completed} to one where \textit{both are} also requires that the activity after the gateway gets activated. State change validation also includes checking write permissions for global variables and contrasts the evaluation of arithmetic expressions with the proposed path for exclusive gateways.

%\paragraph{Variable write permission}
%\label{variable_permission}

%The program must ensure that the global variables can only change in the tasks that have the write permissions for that specific variable. This is done in our custom BPMN attributes described in section \ref{bpmn_attr}.

%\paragraph{Exclusive gateway validation}
%\label{exclusive_assert}
%After the state's task vectors are validated, We need to ensure that the right path was chosen after the start of an exclusive gateway. This is why the arithmetic expression on the chosen edge is evaluated. 

%\paragraph{Message validation}
%\label{message_hash}
Finally, the message-handling validation logic involves two major validation aspects. First, a message hash must be provided when a participant wants to mark a Message Throw event as ''completed''. We assume the actual message to be passed off-chain. Second, when a participant wants to mark a Message Catch event as ''completed'', we must ensure that the corresponding Message Throw event is also marked as completed. The receiver contrasts the message with the hash value; if this fails, we assume that the further steps are either captured in the process logic or the sender and receiver coordinate corrective transmission off-chain. %Contrasting the received message with the hash value has to be done by the receiver; if this fails, we assume that the corresponding steps are either captured in the process logic or the sender and receiver coordinate corrective transmission off-chain.

%\paragraph{Pseudo code}
%\todo{ref} includes detailed pseudo-code for the state change validity check.

%\subsubsection{Authorisation}
%\label{authorisation}
%To authorize a participant, the program proves that the participant has the private key, which corresponds to the task's specified public key. 

%The program also proves that the signature supplied as public input can be decrypted with that public key. The signed message has to be the hash of the previous state (with the randomness) and the hash of the new, proposed state (with a new random number). This signature is then stored in the smart contract.

%An EdDSa implementation is available in the Zokrates standard library. It can verify that a party has a private key with a corresponding public key and verify signatures.

%% file: content/protocol.tex
\subsection{The zkWF protocol}
\label{protocol}
%The zkWF protocol largely follows the general ''proofs over commitments and proposed commitment updates'' pattern customary in blockchain applications of ZKPs (as depicted in Figure \ref{fig:basic_update}). The \textit{process manager smart contract} component is fairly simple: on the one hand, it stores and updates commitments, and on the other hand, it checks ZKPs over the new commitments proposed in incoming blockchain transactions. Specifically, the process manager smart contract stores the following data (using the notations introduced earlier):

The protocol flows through the orchestrator smart contract and is simple in light of the earlier sections. The smart contract state contains the following elements:

%\subsubsection{Process manager smart contract design}
%\label{smartcontract}
%This smart contract contains the logic that is executed on the blockchain. It is responsible for the integrity of the business process execution. It must not contain the current state in plain text. It is designed to be relatively simple since most of the computation is off-chain. Figure \ref{fig:sc_design} shows the overall design of this smart contract.

%The process manager smart contract stores the following data:
\begin{itemize}
	\item $\hcurr =\textit{hash}(\scurr || \rcurr)$ %- the hash of the current state and some randomness
	\item $\Ccurr=\textit{enc}(\scurr,\rcurr)$ %- the ciphertext of the current state and the randomness used in the hashing encrypted with a common, predefined key (each participant should have this key)
	\item $\Scurr=\textit{sig}(h_{prev}||\hcurr)$ %-  the previous states' and the current states' hashes concatenated, signed by the last acting participant.
\end{itemize}

\smallskip

\noindent where $enc$ denotes encryption with the group encryption key and method (see Section \ref{sec:conf}). %and $sig$ denotes signing by the party who proposed the second hash commitment in the concatenation (i.e, $\hcurr$). 
Update request transactions of the smart contract carry the following arguments:

%A participant can update these variables with a function. This function has the following arguments but may have others:
\begin{itemize}
    \item $\hnew=\textit{hash}(\snew || \rnew)$ %- the hash of the new state and some randomness 
    \item $\Cnew=\textit{enc}(\snew,\rnew)$ %- the ciphertext of the new state and the randomness used in the hashing encrypted with a common, predefined key (each participant should have this key)
    \item $\Snew=\textit{sig}(\hcurr||\hnew)$ %- a signature from the last acting participant of the process
    %\item $\proof$ - a valid proof generated by the corresponding zkWF program.
    \item $p(\hcurr,\Snew,\hnew)$
\end{itemize}

\smallskip

\noindent The last argument is a ZKP of the correspondence of $\hcurr$, $\Snew$ and $\hnew$, under the shared zkWF program. The orchestrator smart contract checks the validity of this proof before accepting the smart contract state change carried by the other arguments.

\subsection{Side-channel attack protections}
%Public BPMN models enable side-channel attacks with a potentially high probability of success. 
Public BPMN models facilitate side-channel attacks on confidentiality. Our work until now aimed to ensure that the trace steps of the BPMN finite automaton remain unintelligible to the external observer; however, the number and timings of transitions still carry information. Most BPMN models are simple enough to infer a usable probability distribution of possible states and traces from just these observations.

Constant-time execution and delay randomization are two apparent protection options, though both introduce artificial delays. Consider a constant-time token passing ring schedule with dummy operations as our already established scheme. For $n$ participants, we determine a suitable time quantum $t$ with which it is acceptable to wait for $(n-1)t$ to delay the ''posting'' of any state change. During process execution, at the beginning of the $i$-th epoch, participant $i \mod n$ checks whether it needs to send a state update transaction. If yes, it does; if not, it issues a ''fake update'' transaction. After terminating the process, a long fake update stream is advisable. As long as enough participants meet their fake update obligations and adhere to their epochs, external observers only see a heartbeat-like stream of uninterpretable transactions and can determine even the time of termination only with low probability.

\section{Security properties}
The presented approach addresses the security requirements defined in Section~\ref{subsec:prob} as discussed in this section.
\subsection{Integrity}
Property I1 holds in the sense that we carefully implement a strict subset of BPMN semantics, but we acknowledge that future work should create an explicit proof of conformance. I2 holds due to application-level cryptographic authentication; I3 due to cryptographic authentication and the very simple sub-logic of enabling activities and message operations.
\subsection{Availability}
A1 holds due to I2 and the blockchain fairness assumption -- which is mild for high-throughput public and cross-organizational blockchains. A2 holds only under the disincentive assumption of Section~\ref{sec:design}. However, the assumption is not strong for domains with a credible threat of legal or regulatory action (e.g., finance). A participant can also perform a denial of service attack with a constant stream of malicious fake updates. The disincentive assumption applies here, too, but fake update regimen-dependent defences can also be introduced in the smart contract (e.g., epoch schedule enforcement). A3 holds due to a smart contract accounting for state and trace and the blockchain platform assumptions.
\subsection{Confidentiality}
The C1 guarantee has two layers. At the platform level, all transactions can originate from single-use addresses on pseudonymizing platforms -- e.g., Ethereum. In Hyperledger Fabric, the Identity Mixer protocol suite for transactor anonymization and unlinkability can be used similarly. At the application level, transaction payloads and smart contract states contain only hashed, signed and encrypted data. Hashing is straightforward; for the signed content, note that EdDSA signatures do not provide a way to recover the signer’s public key from the signature or to determine whether the same key was used to sign two different messages. For the encrypted state, if not a single, group-shared secret is used, an application should choose an encryption scheme where the participant keys cannot be recovered.

C2 depends on external data and transaction uninterpretability, which flows from the cryptographic measures, and transaction unlinkability, which also relies on the measures for C1. It also requires sufficient side-channel protection, for which we have at least one strong (not necessarily efficient) option.

%% file: content/implementation.tex
\section{Implementation, testing and performance}
%This section describes how we implemented the zkWF programs and how we integrated the zwWF protocol to our tool.
%\subsection{zkWF implementation}
%This section describes how the zkWF program was implemented in ZoKrates.
%\subsubsection{ZKP framework}
The ZoKrates toolkit is a central component in our framework; the current implementation uses version {0.7.13}\footnote{See \url{https://github.com/Zokrates/ZoKrates/releases}}. ZoKrates was the ZKP toolkit with the best-fitting programming language and ZKP scheme support during our research. %Although this still holds, the state of the art advances rapidly\footnote{A not peer-reviewed, but regularly maintained and curated overview can be found at: \url{https://github.com/matter-labs/awesome-zero-knowledge-proofs}}; we expect that in the future, we will have to reevaluate the available frameworks for our purposes.

%To implement zkWF programs, We chose Zokrates because this seemed like the most advanced solution at the time of writing. It also makes it easy to generate a verifier smart contract which would be tedious to write manually. We used ZoKrates version \href{URL}{0.7.13} at the time. It is the latest version as of writing this.

\subsection{Code generation}
Our code generator, implementing the transformation logic denoted in Figure~\ref{fig:zkwfProgram}, is a custom development in Kotlin. This component generates a zkWF program from an XML-serialized BPMN model, relying on ZoKrates template files. First, the model is encoded, as we outlined earlier; then, it generates the code for the described stages of computation and checks.% code for calculating the state hashes, checking the variable write permissions, ensuring exclusive gateway paths, and verifying message sending. %Distribution and validation of the resulting zkWF program among the process participants are not covered by the framework. 
 We also generate the orchestrator smart contracts for EVM-based blockchains (Solidity version {0.8.0}) and Hyperledger Fabric (Java ''chaincode''). %Solidity smart contracts are derived from the verifier smart contracts ZoKrates generates for zkWF programs; Fabric ones do not have a precursor. In Fabric, smart contracts -- ''chaincodes'' -- run in Docker containers. We created a custom chaincode container, where the chaincode and the ZoKrates toolkit (for proof verification) are bundled together.

%\section{Smart contract implementation}
%To make our approach less dependent on one technology, We implemented the process manager smart contract for two distributed ledger systems: Ethereum and Hyperledger Fabric.
%\subsection{EVM}
%The process manager smart contract (see section \ref{smartcontract}) is implemented in Solidity version 0.8.0. It is derived from a Verifier smart contract generated by ZoKrates.

%\subsection{Hyperledger Fabric}
%\label{fabric_verifier}
%ZoKrates can only generate smart contract verifiers in Solidity. Fortunately,  Hyperledger Fabric chaincodes run in standard docker containers and are written in traditional programming languages like Java. This means we can use the ZoKrates by executing the commands in the container with a ProcessBuilder class.

%The rest of the smart contract implementation is analogue to the Solidity one.

\subsection{Client side}
We created a simple participant-side SDK, which wraps ZoKrates and incorporates the Web3J wallet library. We also created a TornadoFX-based desktop GUI application (''WorkFlow GUI'') for testing and demonstration purposes. The GUI supports all key participant-side actions: monitoring a process manager smart contract for changes, retrieving state, creating process step proposals, computing their witnesses and proofs, and submitting update proposals.

WFGUI also incorporates a process modeller for our BPMN subset and extensions through an embedding of bpmn-js\footnote{See \url{https://bpmn.io/toolkit/bpmn-js/}}; supports testing through preassembled smart contract call sequences; and supports process manager smart contract deployment to Ethereum-based blockchains. A demonstrational video is available in our repository.
%We made a GUI application called WFGUI (WorkFlow GUI) to fully integrate my approach into a tool. We implemented it in Kotlin (as with the rest of the parts) with the TornadoFX library (Kotlin wrapper for JavaFX).

%The GUI itself is separated into three different tabs. One for modelling, one for testing, and the last for deploying and operating.

\subsection{Functional testing}
We assembled a suite of \textit{simple test cases}, %and known important \textit{''corner cases''}
based on the test model suites of the tools cited in Section \ref{sec:procorch}. BPMN model size and complexity influence zkWF program size and complexity, which, in turn, determine proof computation times and on-chain verification costs. To evaluate the practical feasibility of our approach, the leasing model from Section \ref{sec:motiv} was used as our \textit{representative test case}.  %in the practice -- and, while in this paper we have not addressed this question yet, there is a clear path from the current solution towards a set of process manager smart contracts, which collectively manage the state commitments of a process hierarchy and remain efficient from the point of view of proof generation and verification costs.

%The test cases are executed by a custom test scenario runner framework, which has a CLI interface (for CI/CD pipeline integration) in addition to its integration into WFGUI. 

%\begin{figure*}
%  \includegraphics[width=\textwidth,keepaspectratio]{img/leasing.png}
%  \caption{Leasing model}
%\end{figure*}

%To make sure this method of executing business processes is correct, We propose a few ways of challenging it.
%\subsection{Test cases}
%Our first method of testing our approach is by defining test cases and seeing if they can be used in our tool.
%\subsubsection{Simple \& corner cases}
%These test cases were designed to ensure that every supported element works. We also wanted to know how the program reacts to corner cases where the model syntax is correct but the semantics are questionable.
%\subsubsection{Representative test}
%As a Representative test, We wanted to use a complex, real-world example BPMN model. The goal is to see how the program reacts to more oversized state objects (e.g., more executable events, messages, etc.).

%\subsubsection{Testing framework}
%\label{testing_frame}
%To run these tests We designed a testing framework. This framework can run individual test scenarios (i.g. separate steps) and run them as a batch.

%The framework also has a GUI and CLI interface. The GUI can be accessed from WFGUI. The CLI interface is designed to work without user interaction. This makes it easy to integrate as a CI/CD pipeline.

%% file: content/results.tex
\subsection{Performance evaluation}
%In this chapter, We provide quantitative performance figures to demonstrate the feasibility of the zkWF approach.
%\subsection{Hardware used for testing}
In addition to functional testing (compliance with model semantics, proper enforcement of authorization aspects and proper handling of compliant/noncompliant proofs), we used our test suite to evaluate key performance metrics of the approach. Performance tests were performed on a desktop PC (AMD Ryzen 7 2700, 16 GB of DDR4 memory). 

In Ethereum, smart contract execution steps, measured in ''gas'', incur a cryptocurrency cost, paid by the transaction-requesting user. For measurements of gas used, we used a private, one-node Ethereum test network with version {1.10.25} of geth, the official Go implementation of the Ethereum protocol. Blockchain-side efficiency measurements are largely irrelevant for Hyperledger Fabric, which has no ''gas'' notion and where the smart contract execution layer is highly resource-scalable. Table \ref{tab:compare} summarizes the relevant size metrics of our test cases.

%\subsubsection{Software used for testing}
%To run each test scenario, We used our testing framework (section \ref{testing_frame}). To measure the gas required to run on an Ethereum-compatible blockchain, We set up a private test network using geth version 1.10.25.
%\subsubsection{Comparing test cases}
%Table \ref{tab:compare} compares the test cases based on their sizes. These aspects ruffly measure the complexity of each model. In the following sections, We want to find how the complexity of a model is linked to compilation time, setup time(the time it takes to do the trusted setup phase), proof creation time (the amount of time that is needed to produce a proof),  and gas usage on Ethereum.
\begin{table}[H]
    \centering
    \caption{BPMN model and test case characteristics}
    \begin{tabular}{|c|c|c|c|c|c|}
        \hline
        \textbf{Case} & \textbf{Vertices} & \textbf{Edges} & \textbf{Executable} & \textbf{Size of $P$} & \textbf{Scenarios}\\
        \hline
        \hline
        Test 1 & 5 & 4 & 3 & 3 & 3\\
        \hline
        Test 2 & 9 & 10 & 5 & 7 & 9\\
        \hline
        Test 3 & 8 & 8 & 4 & 4 & 4\\
        \hline
        Test 4 & 6 & 5 & 2 & 3 & 2\\
        \hline
        Test 5 & 14 & 12 & 10 & 10 & 10\\
         \hline
        Repr. & 68 & 69 & 50 & 54 & 52\\
         \hline
    \end{tabular}
    \label{tab:compare}
\end{table}
%\subsubsection{Results}
%We ran all tests sequentially, and all test cases were successful. 
Table \ref{tab:timing} summarizes the runtimes of the off-chain computations. Compilation and zk-SNARK setup were executed once; proving time is the sum of computing the witness and generating the proof, and we give an average over the scenarios. The measurements indicate that our approach is practically feasible for real-life models.

\begin{table}[H]
    \centering
    \caption{Off-chain computation runtimes}
    \begin{tabular}{|l|c|c|c|c|}
    \hline
    \textbf{Case} &\textbf{Compilation time} & \textbf{Setup time} &  \textbf{Proving time avg.} \\
    \hline
    \hline
        Test 1 & 27.22 s & 129.58 s & 55.0 s \\
    \hline
        Test 2 & 48.32 s & 182.80 s & 88.67 s \\
    \hline
        Test 3 & 28.55 s & 129.69 s & 53.40 s  \\
    \hline
        Test 4 & 27.14 s & 128.82 s & 53.21 s  \\
    \hline
        Test 5 & 30.74 s & 133.44 s & 54.10 s\\
    \hline 
        Repr. & 81.02 s & 187.33 s & 122.47s \\
    \hline
    \end{tabular}
    \label{tab:timing}
\end{table}
%Table \ref{tab:gas_usage} shows how the smart contract performs if deployed on an Ethereum-compatible blockchain. The table demonstrates how much gas is used to deploy the smart contract (one-time fee). It also describes how much gas is required to update the current state on the blockchain.

Table \ref{tab:gas_usage} summarizes the gas costs of smart contract deployment and smart contract calls in the zkWF protocol. Note that although the representative model is 5-6 times larger than the simple ones, the smart contract call gas cost is only moderately higher. As the hashes, signatures, and proofs have a fixed length, gas usage variability is driven by the size of the encrypted version of the current state. In the measurements, we use state cleartext instead of ciphertext to eliminate the impact of the not-constrained encryption. %(and possibly prior compression).

%It is essential to point out that even though the representative model is 5-6 times larger than the smaller models shown in table \ref{tab:compare}, the gas usage is only $11.90$ per cent higher than the lowest in previous steps.

%This was expected, since the hashes, the signatures, and the proofs have a fixed length. This means the only thing that drives gas usage higher in larger models is the encrypted version of the current state.
\begin{table}[H]
    \centering
    \caption{Gas usage on Ethereum}
    \begin{tabular}{|l|c|c|c|c|}
    \hline
    \textbf{Case} & \textbf{Deployment gas usage} & \textbf{Update gas usage avg.} \\
    \hline
    \hline
        Test 1 &  2,098,786 gas & 490,507 gas \\
    \hline
        Test 2 & 2,098,990 gas & 497,780 gas \\
    \hline
        Test 3 & 2,098,498 gas & 493,705 gas \\
    \hline
        Test 4 & 2,078,071 gas & 503,817 gas \\
    \hline
        Test 5 & 2,161,039 gas & 491,783 gas \\
    \hline 
        Repr. &  2,408,635 gas & $548,898$ gas \\
    \hline
    \end{tabular}
    \label{tab:gas_usage}
\end{table}

%\subsection{Comparison with existing solutions}
%Performance-wise, it is hard to compare our approach to others, since, to our knowledge, we are the first to use zero-knowledge proofs to hide the current state of BPMN execution on-chain. %Despite some common points, even \cite{toots_msc} remains incommensurable due to differences in goals, basic approach and tools. 
Due to the novelty of our approach, it is comparable with the state of the art only in gas costs. Deployment is on par with, or is better than, the existing solutions. However, the cost of updating the state is significantly higher; ChorChain uses about 92,905 gas on average for each message and Caterpillar is similar to ChorChain. 

This ''confidentiality premium'' is certainly not acceptable on the Ethereum mainnet. Still, it can be argued that the high gas price on the mainnet has ''priced out'' all use cases that were not strictly crypto-financial years ago. On the other hand, at the time of this writing, on multiple alternative EVM-based public blockchains, the gas costs of our operations translate to fractions of 1 USD. Additionally, our approach has evident usage potential on purpose-created, permissioned, cross-organizational blockchains; in this case, the gas cost is a technical consideration and low enough to allow for dozens of transactions per block under customary block gas targets. Lastly, we store encrypted state on-chain ''only'' to fulfil requirement A3 the simplest way; highly available off-chain data storage with blockchain-based integrity assurance is a common technique.

%% file: content/limitations.tex
\section{Threats to validity and future work}
\label{sec:futur}
%This section shows the limitations of this approach and the reason behind them.
%Earlier, we addressed our current constraints with respect to BPMN models which are admissible under our current approach; future work will target lifting these limitations. We dealt with the inability of ZoKrates not being able to verify that a given ciphertext is the encrypted form of a given message with a given key by constructing the protocol so that parties submitting a noncompliant ciphertext can always be identified irrepudiably. We accept smart contract gas costs as an outstanding issue; however, it is one which does not truly limit the applicability of our approach on permissioned-consensus platforms and one which we can reasonably expect to become a non-issue for the Ethereum mainnet, too. We will also investigate whether we can sidestep this issue by transforming our approach into a Layer 2 ZKP rollup scheme, where we can, at the very least, amortize gas costs across batches of process manager smart contract updates.

We see compliance with BPMN operational semantics as a non-negligible threat to validity, especially after our planned future extension of the supported BPMN subset. For the approach presented in this paper, we only tested compliant behavior and not formally prove it; this remains future work. %Here we note that as long as all participants are aware of the way BPMN models are translated to admissible and nonadmissible state changes in zkWF programs, even divergences from the standard-prescribed semantics may be acceptable, but this is clearly not a fully satisfactory answer.

Impractical proof time for much larger BPMN models is also a threat. We plan to introduce the capability to handle \textit{hierarchical} process models. We expect that we can instantiate orchestrator smart contracts for sub-processes in a way that coordinates the commitment-management across the levels, but controls proof obligation complexity by requiring proof generation only for a limited-size model part for each update.

While the ring schedule ''fake updates'' approach is evidently correct for adhering participants (and, we surmise, for mostly adhering participants), side-channel protections is an open line of research. We plan to analyse the ring schedule scheme under various participant failure models and compare it with delay randomization schemes. Metrics for measuring the guaranteed level of protection through fake updates are necessary, too. Differential privacy metrics worked out for publicly observable messaging settings with a ''hide-in-the-crowd'' approach similar to ours \cite{seres_effect_2022} promise to be adaptable.

Lastly, we note that there are stronger versions of our collaboration confidentiality model through additional \textit{inter-collaborator confidentiality constraints}; it is an interesting question how our approach can be extended to such settings.

%Formally proving the operational semantics-preserving nature of our transformation logic would be a possible approach, but that would first need migrating the implementation to a transformation model where the translation rules themselves are first-class objects (such as in the model transformation platform VIATRA \cite{10.1007/978-3-319-21155-8_8}), and even then, it would be a highly complicated endeavour. 

%Instead, future work will first investigate proving behavioural equivalence between reference BPMN behaviour and zkWF programs on a \textit{model-by-model basis}, as a prerequisite check integrated into the toolchain. Recent work has formalized BPMN collaboration semantics in a way amenable for state space model checking \cite{CORRADINI2021111007}; this opens up the possibility to perform bisimulation analyses between the state transitions of models under ''authoritative'' semantics and under zkWF program encoded semantics. This approach will have the added benefit that it facilitates the introduction of property checking (soundness, safeness and application-specific requirements) on the BPMN models themselves.

%% file: content/conclusion.tex
\section{Conclusion}
In this paper, we presented a collaboration confidentiality-preserving approach for the smart contract-based orchestration of business collaborations, captured as BPMN 2.0 models. Our protocol is a novel, and to our knowledge, first-of-its-kind solution, which we validated functionally as well as evaluated from the resource usage and gas cost points of view. We also described a full toolchain prototype which we made available as open-source software.

%Currently, the end-to-end toolchain has full integration only for Ethereum; Hyperledger Fabric integration is partial. We will finish the integration with Fabric and also investigate whether we can also support emerging ''blockchain middleware'' technologies, such as Hyperledger Firefly, with the same toolset.

%As we noted, the gas usage of the process manager smart contracts is not satisfactorily low for all deployment options. In addition to potentially utilizing novel developments in Ethereum platform technology, 

%We would also love to reduce the gas usage of this approach on Ethereum to make transactions cheaper. This would be possible by generating proofs for making several steps in a batch.

%Last but not least, there is further work to be done on guaranteeing BPMN operational semantics compliance.

%% file: main.bbl
\begin{thebibliography}{10}
\providecommand{\url}[1]{#1}
\csname url@samestyle\endcsname
\providecommand{\newblock}{\relax}
\providecommand{\bibinfo}[2]{#2}
\providecommand{\BIBentrySTDinterwordspacing}{\spaceskip=0pt\relax}
\providecommand{\BIBentryALTinterwordstretchfactor}{4}
\providecommand{\BIBentryALTinterwordspacing}{\spaceskip=\fontdimen2\font plus
\BIBentryALTinterwordstretchfactor\fontdimen3\font minus \fontdimen4\font\relax}
\providecommand{\BIBforeignlanguage}[2]{{%
\expandafter\ifx\csname l@#1\endcsname\relax
\typeout{** WARNING: IEEEtran.bst: No hyphenation pattern has been}%
\typeout{** loaded for the language `#1'. Using the pattern for}%
\typeout{** the default language instead.}%
\else
\language=\csname l@#1\endcsname
\fi
#2}}
\providecommand{\BIBdecl}{\relax}
\BIBdecl

\bibitem{van2016business}
W.~M. Van Der~Aalst, M.~La~Rosa, and F.~M. Santoro, ``{Business process management: Don't forget to improve the process!}'' \emph{Business \& Information Systems Engineering}, vol.~58, no.~1, pp. 1--6, 2016. doi: 10.1007/s12599-015-0409-x

\bibitem{POURMIRZA201743}
S.~Pourmirza, S.~Peters, R.~Dijkman, and P.~Grefen, ``{A systematic literature review on the architecture of business process management systems},'' \emph{Information Systems}, vol.~66, pp. 43--58, 2017. doi: 10.1016/j.is.2017.01.007

\bibitem{bpmn}
\BIBentryALTinterwordspacing
{Object Management Group}, ``\BIBforeignlanguage{en}{{Business Process Model and Notation (BPMN), Version 2.0}}.'' [Online]. Available: \url{https://www.omg.org/spec/BPMN/2.0/}
\BIBentrySTDinterwordspacing

\bibitem{CHINOSI2012124}
M.~Chinosi and A.~Trombetta, ``{BPMN: An introduction to the standard},'' \emph{Computer Standards \& Interfaces}, vol.~34, no.~1, pp. 124--134, 2012. doi: 10.1016/j.csi.2011.06.002

\bibitem{10.1145/3183367}
J.~Mendling \emph{et~al.}, ``{Blockchains for Business Process Management - Challenges and Opportunities},'' \emph{ACM Trans. Manage. Inf. Syst.}, vol.~9, no.~1, 2018. doi: 10.1145/3183367

\bibitem{zokrates}
J.~Eberhardt and S.~Tai, ``{ZoKrates - Scalable Privacy-Preserving Off-Chain Computations},'' in \emph{2018 IEEE International Conference on Internet of Things (iThings) and IEEE Green Computing and Communications (GreenCom) and IEEE Cyber, Physical and Social Computing (CPSCom) and IEEE Smart Data (SmartData)}, 2018. doi: 10.1109/Cybermatics\_2018.2018.00199 p. 1084–1091.

\bibitem{cachin2017blockchain}
\BIBentryALTinterwordspacing
C.~Cachin and M.~Vukoli{\'c}, ``Blockchain consensus protocols in the wild,'' in \emph{{31st International Symposium on Distributed Computing (DISC 2017)}}, 2017. [Online]. Available: \url{https://drops.dagstuhl.de/opus/volltexte/2017/8016/pdf/LIPIcs-DISC-2017-1.pdf}
\BIBentrySTDinterwordspacing

\bibitem{gorton2006essential}
I.~Gorton, \emph{{Essential Software Architecture}}, 2nd~ed.\hskip 1em plus 0.5em minus 0.4em\relax Springer Berlin, Heidelberg, 2011.

\bibitem{model_driven}
Y.~Ait~Hsain, N.~Laaz, and S.~Mbarki, ``\BIBforeignlanguage{en}{{Ethereum’s Smart Contracts Construction and Development using Model Driven Engineering Technologies: a Review}},'' \emph{\BIBforeignlanguage{en}{Procedia Computer Science}}, vol. 184, p. 785–790, 2021. doi: 10.1016/j.procs.2021.03.097

\bibitem{caterpillar}
O.~López-Pintado, L.~García-Bañuelos, M.~Dumas, I.~Weber, and A.~Ponomarev, ``\BIBforeignlanguage{en}{Caterpillar: A business process execution engine on the ethereum blockchain},'' \emph{\BIBforeignlanguage{en}{Software: Practice and Experience}}, vol.~49, no.~7, p. 1162–1193, 2019. doi: 10.1002/spe.2702

\bibitem{Mercenne_Brousmiche_Hamida_2018}
L.~Mercenne, K.-L. Brousmiche, and E.~B. Hamida, ``{Blockchain Studio: A Role-Based Business Workflows Management System},'' in \emph{2018 IEEE 9th Annual Information Technology, Electronics and Mobile Communication Conference (IEMCON)}, 2018. doi: 10.1109/IEMCON.2018.8614879 p. 1215–1220.

\bibitem{Abid_Cheikhrouhou_Jmaiel_2020}
A.~Abid, S.~Cheikhrouhou, and M.~Jmaiel, ``\BIBforeignlanguage{en}{{Modelling and Executing Time-Aware Processes in Trustless Blockchain Environment}},'' in \emph{\BIBforeignlanguage{en}{Risks and Security of Internet and Systems}}, ser. LNCS, 2020. doi: 10.1007/978-3-030-41568-6\_21 p. 325–341.

\bibitem{lorikeet}
Q.~Lu \emph{et~al.}, ``\BIBforeignlanguage{en}{{Integrated model-driven engineering of blockchain applications for business processes and asset management}},'' \emph{\BIBforeignlanguage{en}{Software: Practice and Experience}}, vol.~51, no.~5, p. 1059–1079, 2021. doi: 10.1002/spe.2931

\bibitem{chorchain}
F.~Corradini \emph{et~al.}, ``\BIBforeignlanguage{en}{{ChorChain: A Model-Driven Framework for Choreography-Based Systems Using Blockchain}},'' in \emph{\BIBforeignlanguage{en}{{Proc. of the 1st Italian Forum on Business Process Management (ITBPM)}}}, 2021, pp. 26--32.

\bibitem{multichain}
------, ``\BIBforeignlanguage{en}{{Model-driven engineering for multi-party business processes on multiple blockchains}},'' \emph{\BIBforeignlanguage{en}{Blockchain: Research and Applications}}, vol.~2, no.~3, p. 100018, 2021. doi: 10.1016/j.bcra.2021.100018

\bibitem{flexchain}
------, ``{Flexible Execution of Multi-Party Business Processes on Blockchain},'' in \emph{2022 IEEE/ACM 5th International Workshop on Emerging Trends in Software Engineering for Blockchain (WETSEB)}, 2022. doi: 10.1145/3528226.3528369 p. 25–32.

\bibitem{hyperledger_fabric}
E.~Androulaki \emph{et~al.}, ``\BIBforeignlanguage{en}{{Hyperledger Fabric: a distributed operating system for permissioned blockchains}},'' in \emph{\BIBforeignlanguage{en}{Proceedings of the Thirteenth EuroSys Conference}}, 2018. doi: 10.1145/3190508.3190538 p. 1–15.

\bibitem{ZKProofCommunity}
\BIBentryALTinterwordspacing
{ZKProof Community}, ``\BIBforeignlanguage{en}{{ZKProof Community Reference}},'' 2022. [Online]. Available: \url{https://docs.zkproof.org/reference.pdf}
\BIBentrySTDinterwordspacing

\bibitem{Groth_2016}
J.~Groth, ``{On the Size of Pairing-based Non-interactive Arguments},'' in \emph{Advances in Cryptology -- EUROCRYPT 2016: 35th Annual International Conference on the Theory and Applications of Cryptographic Techniques}, 2016. doi: 10.1007/978-3-662-49896-5\_11 pp. 305--326.

\bibitem{Groth_Maller_2017}
J.~Groth and M.~Maller, ``{Snarky signatures: Minimal signatures of knowledge from simulation-extractable SNARKs},'' in \emph{{Advances in Cryptology -- CRYPTO 2017: 37th Annual International Cryptology Conference}}, 2017. doi: 10.1007/978-3-319-63715-0\_20 pp. 581--612.

\bibitem{cryptoeprint:2019/1047}
\BIBentryALTinterwordspacing
A.~Chiesa, Y.~Hu, M.~Maller, P.~Mishra, P.~Vesely, and N.~Ward, ``{Marlin: Preprocessing zkSNARKs with Universal and Updatable SRS},'' Cryptology ePrint Archive, Paper 2019/1047, 2019. [Online]. Available: \url{https://eprint.iacr.org/2019/1047}
\BIBentrySTDinterwordspacing

\bibitem{zkws4}
D.~Bennaroch, M.~Campanelli, D.~Fiore, J.~Kim, J.~Lee, H.~Oh, and A.~Querol, ``{Proposal: Commit-and-Prove Zero-Knowledge Proof Systems and Extensions},'' \url{https://docs.zkproof.org/standards/proposals}, presented at the 4th workshop of the ZKProof Community, 19-29 April 2021, online.

\bibitem{9520375}
X.~Sun, F.~R. Yu, P.~Zhang, Z.~Sun, W.~Xie, and X.~Peng, ``{A Survey on Zero-Knowledge Proof in Blockchain},'' \emph{IEEE Network}, vol.~35, no.~4, pp. 198--205, 2021. doi: 10.1109/MNET.011.2000473

\bibitem{9300214}
J.~Partala, T.~H. Nguyen, and S.~Pirttikangas, ``Non-interactive zero-knowledge for blockchain: A survey,'' \emph{IEEE Access}, vol.~8, pp. 227\,945--227\,961, 2020. doi: 10.1109/ACCESS.2020.3046025

\bibitem{Muehlen2013}
M.~z. Muehlen and J.~Recker, ``{How Much Language Is Enough? Theoretical and Practical Use of the Business Process Modeling Notation},'' in \emph{Seminal Contributions to Information Systems Engineering: 25 Years of CAiSE}.\hskip 1em plus 0.5em minus 0.4em\relax Berlin, Heidelberg: Springer Berlin Heidelberg, 2013. doi: 10.1007/978-3-642-36926-1\_35 pp. 429--443.

\bibitem{toots_msc}
T.~Aivo, ``\BIBforeignlanguage{en}{{Zero-Knowledge Proofs for Business Processes}},'' Master's thesis, Univ. of Tartu, 2020.

\bibitem{cryptoeprint:2019/1270}
\BIBentryALTinterwordspacing
J.~Lee, J.~Choi, J.~Kim, and H.~Oh, ``{SAVER: SNARK-friendly, Additively-homomorphic, and Verifiable Encryption and decryption with Rerandomization},'' Cryptology ePrint Archive, Paper 2019/1270, 2019. [Online]. Available: \url{https://eprint.iacr.org/2019/1270}
\BIBentrySTDinterwordspacing

\bibitem{seres_effect_2022}
I.~A. Seres, B.~Pejó, and P.~Burcsi, ``\BIBforeignlanguage{en}{The {Effect} of {False} {Positives}: {Why} {Fuzzy} {Message} {Detection} {Leads} to {Fuzzy} {Privacy} {Guarantees}?}'' in \emph{\BIBforeignlanguage{en}{Financial {Cryptography} and {Data} {Security}}}, ser. {LNCS}.\hskip 1em plus 0.5em minus 0.4em\relax Cham: Springer International Publishing, 2022. doi: 10.1007/978-3-031-18283-9\_7 pp. 123--148.

\end{thebibliography}
